\documentclass[12pt]{article}
\usepackage{amsfonts}
\usepackage{amsmath}
\usepackage{fullpage}
\usepackage{graphicx}
\usepackage{url}

\begin{document}


\title{
    {\bfseries\boldmath Reflections on the Energy of Black Holes}
    \footnote{Essay written for the Gravity Research Foundation 2019
	Awards for Essays on Gravitation.}
}

\author{
	Tevian Dray
	\footnote{Corresponding author.}
	\\[-2.5pt]
	\normalsize
	\textit{Department of Mathematics, Oregon State University,
		Corvallis, OR  97331, USA} \\[-2.5pt]
	\normalsize
	\texttt{tevian@math.oregonstate.edu} \\
	\and
	Carlo Rovelli \\[-2.5pt]
	\normalsize
	\textit{CPT, Aix-Marseille Universit\'e, Universit\'e de Toulon,
		CNRS, F-13288 Marseille, FRANCE} \\[-2.5pt]
	\normalsize
	\texttt{rovelli@cpt.univ-mrs.fr}
}

\date{\normalsize \today}

\maketitle

\begin{abstract}
Inside a black hole, there is no \emph{local} way to say which side of a
sphere is the inside, and which is the outside.  One can easily be gulled by
this fact into mixing up the sign of the energy.  We lead the reader astray
with a na\"\i ve treatment of the energy of a null shell in black hole
spacetimes.  We then resolve the confusion, showing that global, rather than
local, considerations offer good guidance.
\end{abstract}

\newpage


\section{The problem}

\begin{figure}[h]
\centering
\includegraphics[width=2.5cm]{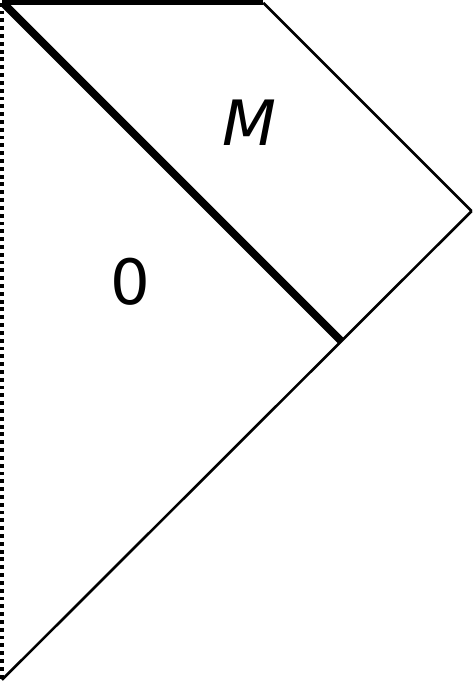}
\caption{A shell of massless matter in flat space collapsing to form a black
hole.}
\label{0M}
\end{figure}

What is the total energy associated with a black hole spacetime?  Surely, the
answer is the parameter $M$ in the line element.  For concreteness, let's
consider the Schwarzschild geometry.  We can model a physical black hole by
collapsing a spherically symmetric shell in flat space.  For simplicity, we
follow Dray and 't Hooft~\cite{CMP}, and join Minkowski and Schwarzschild
spacetimes along a lightlike, spherically symmetric worldtube, as shown in
Figure~\ref{0M}.

The total energy at any instant of time is given by integrating the energy
density over a spacelike hypersurface.  The distributional energy density on
our shell is clearly $M$, the mass of the resulting black hole.

\begin{figure}[h]
\centering
\includegraphics[width=4cm]{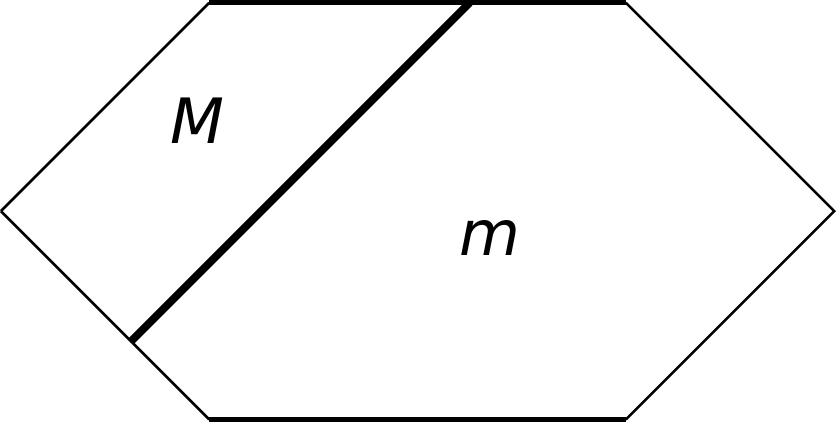}
\caption{A shell of massless matter joining two black holes.}
\label{mM}
\end{figure}

Dray and 't Hooft~\cite{CMP} also considered the similar situation of a shell
collapsing in a black hole background, that is, joining two Schwarzschild
spacetimes of different masses.  As shown in Figure~\ref{mM}, the result is
similar: the energy density of the shell is precisely the difference between
the masses of the two Schwarzschild regions.  Which way do the signs go?  With
the regions labeled as in Figure~\ref{mM}, that is, with mass $m$ ``before''
the shell and mass $M$ ``after'', the shell has energy density $M-m$.

\begin{figure}
\centering
\includegraphics[width=4cm]{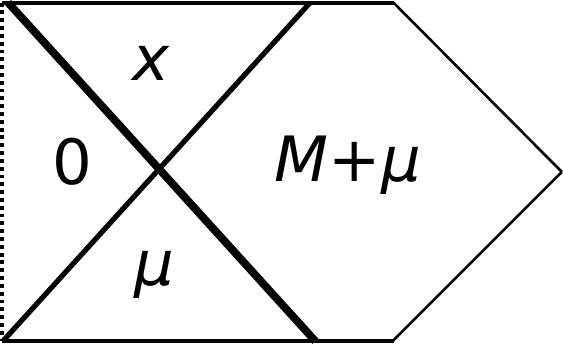}
\caption{A light, outgoing shell crossing a heavier, ingoing shell.}
\label{0mM}
\end{figure}

A somewhat more physical version of the scenario in Figure~\ref{mM} can be
obtained by colliding two shells, replacing the second asymptotic region by
the (flat) interior of a collapsing shell.  Such a scenario is shown in
Figure~\ref{0mM}, in which four spacetime regions are joined across two
crossing shells.  Before the crossing, there is an expanding shell of mass
$\mu$ and a collapsing shell of mass $M$.  Dray and 't Hooft~\cite{CMP}
pointed out that for such shells there is an additional regularity condition
at the crossing point, which here becomes
\begin{equation}
r\bigl(r-2(M+\mu)\bigr) = (r-2\mu)(r-2x) ,
\label{DtH}
\end{equation}
where $r$ is the Schwarzschild radius of the crossing point and $x$ the mass
of the Schwarzschild between the shells after the crossing.  For the causal
structure shown, $2\mu <r<2(M+\mu )$.
\footnote{It is also possible to have $r>2(M+\mu )$, but the causal structure
is then different.}

What is the energy in the two shells after the crossing?  It is
straightforward to argue that the the upper-left shell carries energy $x$, the
mass of the final black hole.  Comparing Figure~\ref{mM} with the
corresponding region in Figure~\ref{0mM} -- with which it must be locally
isomorphic -- leads to the conclusion that the energy of the upper-right shell
must be $x-(M+\mu)$.  Solving for $x$ in Equation~(\ref{DtH}) leads to
\begin{equation}
x = \frac{M}{1-\frac{2\mu}{r}}  ,
\end{equation}
so that
\begin{equation}
x-(M+\mu)
  = -\mu \left(\frac{1-\frac{2(M+\mu)}{r}}{1-\frac{2\mu}{r}}\right)
  > 0  .
\end{equation}
Thus, all four shells carry positive energy.

Wait a minute!  Energy should be conserved.  The total energy at late times
should match that at early times -- the regularity condition at the crossing
point ought to ensure that there is no ``extra'' energy associated with the
crossing.  But if all four shells carry positive energy, then the total energy
at late times is $x+\bigl(x-(M+\mu)\bigr)=2x-(M+\mu)$ which differs from the
energy at early times, which is $M+\mu$.  On the other hand, the
\emph{difference} in shell energies at late times is
$x-\bigl(x-(M+\mu)\bigr)$, which does equal the energy at early times.  Thus,
energy conservation implies that the upper-right shell must have
\emph{negative} energy density.

So which is it?  Is the shell energy positive or negative?  Local
considerations appear to require positive energy, to match Figure~\ref{mM};
global considerations appear to require negative energy, to restore energy
conservation.

\section{Resolution}

A clue can be found by comparing the collapsing shell model (Figure~\ref{0M})
with the usual Kruskal extension of the Schwarzschild geometry, with two
asymptotic regions.  For the collapsing shell, Gauss's Law tells us that, on
any hypersurface, the total energy \emph{inside} any sphere equals the
gravitational flux through the sphere.  This is essentially the mechanism used
to compute the ADM mass~\cite{ADM} at spatial infinity: Compute the flux
through a large sphere, then take the limit to infinity.  Thus, the total
energy on any such hypersurface is $M$, as already claimed.  However, for the
Kruskal extension, the boundary consists of \emph{two} spheres, one in each
asymptotic region.  There is no collapsing shell, and hence no energy, on the
hypersurface between them.  Thus, the total energy on any hypersurface must be
zero.

Furthermore, since the asymptotic region in Figure~\ref{0M} is isomorphic to
one side of the Kruskal extension, the integral over one sphere must be $M$.
The integral over the other sphere must therefore be $-M$.

What's going on here?  It is tempting to argue that this negative energy on
one side is due to the opposite time orientations of the two asymptotic
regions, as manifested for instance by examining $t=\hbox{constant}$ surfaces,
where $t$ is the Schwarzschild time coordinate.  But this argument is clearly
insufficient.  There is a global time orientation in all of these scenarios,
namely up the page.  Interchanging the asymptotic regions by horizontal
reflection can not change the local physics.  Similarly, the local isomorphism
between Figure~\ref{mM} and the corresponding region in Figure~\ref{0mM} is
not affected by horizontal reflection.

However, the \emph{spatial} orientations of the two asymptotic regions are
geometric -- and do not agree.  These opposite spatial orientations clearly
suffice to explain the negativity of the computed energy in one region, since
an \emph{outward} flux in one region corresponds to an \emph{inward} flux
in the other.  Put differently, the spatial orientation is not invariant under
reflection.

Crucially, there is no canonical spatial orientation defined locally inside
the black hole.  Don't forget, ``decreasing radius'' has become \emph{time}!
Inside a black hole there is no local sense in which a sphere has a naturally
defined inside and outside!

This, then, is the resolution of the apparent paradox: Figure~\ref{mM} tells
us that the energy of the shell is positive \emph{as seen in the asymptotic
universe in which the shell originated}, that is, using the spatial
orientation inherited from that region.  But that region does not exist in
Figure~\ref{0mM}; as seen from the only asymptotic region we have, the shell
energy is negative.  Energy conservation is restored.
Similar arguments can be used to show that global energy conservation holds in
all of the colliding shell models in~\cite{CMP}, including those with two
asymptotic regions.

The moral: Always look both ways before crossing!

\vspace{-5pt}
\section*{Acknowledgments}

This essay grew out of a simple question asked by one of us (CR) of the other
(TD), which however resulted in a flurry of 50 email messages over the
subsequent week, mostly expressing strong disagreement, until finally
converging on a resolution


\vspace{-5pt}
\begin{thebibliography}{9}

\bibitem{CMP}
Tevian Dray and Gerard 't Hooft,
{\it The Effect of Spherical Shells of Matter on the
 Schwarz\-schild Black Hole},
Commun.\ Math.\ Phys.\ {\bf 99}, 613--625 (1985).

\bibitem{ADM}
Arnowitt R., Deser S., and Misner C. W.,
{\it Canonical Variables for General Relativity},
Phys.\ Rev.\ {\bf 117}, 1595--1602 (1960).

\end{thebibliography}
\end{document}